\documentstyle[amssymb,aps,preprint]{revtex}
\def\be{\begin{equation}}
\def\bea{\begin{eqnarray}}
\def\bma{\begin{mathletters}}
\def\ee{\end{equation}}
\def\eea{\end{eqnarray}}
\def\ema{\end{mathletters}}

\begin{document}
\author{Vlatko Vedral\footnote{e-mail address: vlatko.vedral@qubit.org}}
\title{On bound entanglement assisted distillation }
\address{Centre for Quantum Computation, Clarendon Laboratory, University of Oxford,\\
Parks Road OX1 3PU}
\date{\today}
\maketitle

\begin{abstract}
We investigate asymptotic distillation of entanglement in the presence of an
unlimited amount of bound entanglement for bi-partite systems. We show that
the distillability is still bounded by the relative entropy of entanglement.
This offers a strong support to the fact that bound entanglement does not
improve distillation of entanglement.
\end{abstract}
\vspace*{1cm}
PACS number(s): 03.65.Bz, 89.70.+c,89.80.+h\\
{\bf keywords}: entanglement, relative entropy
\vspace*{1cm}

\bigskip

\noindent Recent years have witnessed an explosion of interest in
understanding and quantifying the amount of entanglement in a given state of
bi-partite and multi-partite quantum systems (for a review see \cite{Plenio1}%
). Entanglement of bi-partite systems in a pure state is well understood
asymptotically; it can be quantified by a single measure - von Neumann's
reduced entropy \cite{Bennett1},\cite{Bennett2},\cite{Vedral1}. In the
finite (non-asymptotic) case \cite{Lo}, however, there are still many
surprises. For example, one of these surprises is the existence of catalytic
processes recently found by Jonathan and Plenio \cite{Jonathan}.
Multi-partite systems are, on the other hand, not well understood at all;
even the bi-partite mixed states display a number of counter-intuitive
properties such as the existence of bound entanglement \cite{Horodecki1}.
Also, it is not known with what maximal efficiency can entanglement be
distilled from a general bi-partite mixed state. Present work lies in the
domain of asymptotic distillation of bi-partite quantum systems and
investigates the usefulness of bound entanglement in this process. It is
known that there are three different types of entanglement for bi-partite
systems \cite{Horodecki1} and they are: 1. free entangled states (FE), i.e. 
states from which pure singlets can be distilled by local operations aided with classical
communication (LOCC); 2. bound entangled states (BE), which are inseparable,
but cannot be distilled by LOCC and 3. separable states, which contain no
entanglement and consequently cannot be distilled either. There are several
bounds for the yield of distillation (discussed below), the number which is sometimes also
called the entanglement of distillation \cite{Bennett2}. In this paper we
consider the following problem (first posed by the family Horodecki in \cite
{Horodecki1}). Suppose that we have an unlimited supply of bound
entanglement in addition to LOCC. Can this help in distilling more singlets
then we could have without the presence of bound entanglement? We provide an upper bound
to the bound-entanglement assisted distillation which strongly suggests that
the answer is no. If, in addition, the entanglement of formation for BE states was
strictly non-zero (which would be true if the entanglement 
of formation was additive \cite{Bennett2}, but otherwise is not known), 
this would then make them more than useless: they 
would require a certain amount of entanglement to be created by local means even
asymptotically, but would not be able to enhance distillation in any way. 

\noindent

\bigskip

\noindent A distillation procedure\noindent\ involves two parties, Alice and
Bob, who share a certain number of bi-partite systems in some quantum state.
Their task is to convert this into as large a number of singlets (or any other
maximally entangled state of two qubits) as
possible, by acting only locally, i.e. on their own systems, and
communicating classically results of their actions to each other. 
This is a very important protocol since we know that maximally
entangled pairs achieve a higher fidelity of quantum information
transmission and processing in general. Quantum relative entropy (defined
below) is a quantity that is non-increasing under a general quantum
evolution \cite{Lindblad1} and in particular under LOCC. It can therefore
help us to produce an upper bound on the amount of distillable entanglement 
\cite{Vedral2}. We will show that the number of singlets (or any other
maximally entangled two qubit states) that can be distilled from $n$ copies
of $\sigma $ is bounded from the above by $\min_{\omega \in {\cal B}%
}S(\sigma ^{\otimes n}||\omega ),$ where $\omega $ is of the same
dimensionality as $\sigma ^{\otimes n}$, $S(\sigma ||\omega ):=tr\left\{
\sigma \log \sigma -\sigma \log \omega \right\} $ is the quantum relative
entropy between $\sigma $ and $\omega$ and ${\cal B}$ is the set of BE and
separable (disentangled) states together. 
In other words, we will show that
the entanglement of distillation $D(\sigma )\leq \frac{1}{n}\min_{\omega \in 
{\cal B}}S(\sigma ^{\otimes n}||\omega ).$ This then provides an upper bound
on distillation, which is achievable in distillation of pure states \cite{Bennett1} and Bell-diagonal states with only two no-zero eigenvalues \cite{Bennett2}. 
A similar bound and this general method of providing a
bound was first proposed by Vedral et. al. \cite{Vedral1} and Vedral and
Plenio \cite{Vedral2}, who performed the minimisation over separable states. This was 
then recently improved by Rains \cite{Rains1},\cite{Rains2} by introducing a minimisation
over BE states in addition to separable states and enlarging the set of operations that Alice and Bob can perform. Our main result will then be to show that this bound cannot be improved by
adding any amount of bound entanglement to aid the distillation.

\bigskip

\bigskip

\noindent Before we present the main result we explain our notation. The
symbol $\psi ^{\otimes n}$ means $n$ copies of the state $\psi .$ On the
other hand, $\omega _{m}$ means that $\omega $ lives in the space of $m$
bi-partite systems, but is not necessarily of the product nature. The dimensionality of these systems will always be clear from the problem we are considering. In this case the
subscript will be omitted when there is no chance of confusion. $\Lambda $
will denote a LOCC. Greek symbols $\omega ,$ $\beta $ and $\gamma $ will
denote bound entangled states. $\psi $ always denotes a maximally entangled
state of two qubits, and $\sigma $ will be the state from which we wish to
distill some entanglement and could be free entangled, bound entangled or
separable (this is a bi-partite system of any dimensionality). $D$ will
stand for the (unassisted) entanglement of distillation and $D^{BE}$ for the
bound-entanglement assisted entanglement of distillation, where we allow an
unlimited amount of bound entanglement.

\bigskip

\noindent {\bf Lemma 1}. If $\psi $ is any maximally entangled state of two
qubits, then $\min_{\omega \in {\cal B}}S(\psi ^{\otimes n}||\omega )=n.$
(The state $\omega $ lives in the Hilbert space of dimension $2^n\times 2^{n}$)$.$

\noindent {\it Proof:} By definition of the relative entropy we have that $%
\min_{\omega \in {\cal B}}S(\psi ^{\otimes n}||\omega )=min_{\omega \in 
{\cal B}}-\langle \psi ^{\otimes n}|\log \omega |\psi ^{\otimes n}\rangle $.
But, the logarithmic function is concave so that 
\[
\min_{\omega \in {\cal B}} -\langle \psi ^{\otimes n}|\log \omega |\psi ^{\otimes n}\rangle \geq
\min_{\omega \in {\cal B}} -\log \langle \psi ^{\otimes n}|\omega |\psi ^{\otimes n}\rangle 
\]

\noindent However, according to the recent result of the family Horodecki 
\cite{Horodecki3}, since $\omega $ is a bound entangled state, then its
fidelity with the maximally entangled state cannot be larger than the
inverse of the half dimension of that state, so that $\langle \psi ^{\otimes n}|\omega |\psi
^{\otimes n}\rangle \leq 1/2^{n}.$ Thus, 
\[
\min_{\omega \in {\cal B}}S(\psi ^{\otimes n}||\omega )\geq -\log (1/2^{n})=n 
\]

\noindent But we know that this minimum is achievable by the (in fact
separable) state $\omega =\rho ^{\otimes n},$ where $\rho $ is obtained from 
$\psi $ by removing the off-diagonal elements in the Schmidt basis.
This therefore completes our proof ${}_{\Box}$.

\bigskip

\noindent This lemma is very important since it tells us that the relative
entropy can be used to produce an upper bound on distillable entanglement.
Namely, if we are starting with $n$ copies of state $\sigma ,$ and obtaining 
$m$ copies of $\psi $ by LOCC, then 
\[
D=\frac{m}{n}=\frac{1}{n}\min_{\omega \in {\cal B}}S(\psi ^{\otimes
m}||\omega )\leq \frac{1}{n}\min_{\omega \in {\cal B}}S(\sigma ^{\otimes
n}||\omega ) 
\]

\noindent where the equality follows from lemma 1 and the inequality from
the fact that the relative entropy is non-increasing under LOCC (strictly
speaking, $D=\lim_{n\rightarrow \infty }\frac{m}{n}$ and, of course, $m$ is a
function of $n$, $m=m(n)$ \cite{Rains1}. We will omit limits for two reasons: 1. we wish to keep the notation as simple as possible and 2. all our results apply in the finite case as well). 
The above relies on the fact that
the local operations acting on a bound entangled state cannot produce free
entanglement (this follows trivially from the definition of bound
entanglement). Note also that this achieves a tighter bound than previously
proposed by Vedral and Plenio \cite{Vedral2}, since $\min_{\omega \in {\cal B%
}}S(\sigma ^{\otimes n}||\omega )/n\leq \min_{\omega \in {\cal B}}S(\sigma
||\omega ):=E_{R}(\sigma ),$ where $E_{R}(\sigma )$ is the relative entropy
of entanglement of $\sigma .$ As we have noted, this is closely related to the bound recently
considered by Rains \cite{Rains2}. 

Before we prove the main result we state the assumption under which it is valid.
We assume that if we have two BE states $\beta_1$ and $\beta_2$, then $\beta_1\otimes\beta_2$
is also a BE state. This is true for all the BE states which have a positive partial 
transposition \cite{Horodecki1}, and only these have been known to exist so far \cite{Bennett3,Terhal1}. However, it could happen that a BE state has a negative partial
transposition. In this case, a direct product of two such states might result in a state with free entanglement (this, of course, cannot be proven since we do not know whether a negative partial transposition BE states exist in the first place!). Furthermore, if this 
happens than it is even questionable whether these should
be called BE states to start with. Thus we can safely assume that a direct product of two BE states is also a BE state. We are now ready to state the main result of this paper:

\bigskip

\noindent {\bf Theorem 1.} Suppose that the distillation of $\sigma
^{\otimes n}$ is aided by an unlimited amount of bound entanglement $\beta
^{\otimes k}$ (where $k$ can be any number). Then the distillation yield is
bounded from the above by $\frac{1}{n}\min_{\omega \in {\cal B}}S(\sigma
^{\otimes n}||\omega ).$

\noindent\ {\it Proof:} The proof follows from a chain of inequalities: 
\begin{eqnarray*}
D^{BE} &=&\frac{m}{n}=\frac{1}{n}\min_{\omega \in {\cal B}}S(\psi ^{\otimes
m}||\omega _{m}) \\
&\leq &\frac{1}{n}S(\psi ^{\otimes m}||tr_{k+n-m}\Lambda (\omega _{n}\otimes
\omega _{k})) \\
&\leq &\frac{1}{n}S(\psi ^{\otimes m}\otimes \gamma _{k+n-m}||\Lambda
(\omega _{n}\otimes \omega _{k})) \\
&\leq &\frac{1}{n}S(\sigma ^{\otimes n}\otimes \beta ^{\otimes k}||\omega
_{n}\otimes \omega _{k}) \\
&=&\frac{1}{n}S(\sigma ^{\otimes n}||\omega _{n})+\frac{1}{n}S(\beta
^{\otimes k}||\omega _{k})
\end{eqnarray*}

\noindent The second equality follows from lemma. The second inequality
follows from the fact that partial tracing does not increase the relative
entropy. The third inequality exists because LOCC (in fact, any quantum
operations) cannot increase the quantum relative entropy. The last equality
follows from the definition of the quantum relative entropy.

\noindent But now we arrange to have that $\omega _{k}=\beta ^{\otimes k}$,
so that the second term reduces to zero. Also, we choose $\omega _{n}$ such
that $\min_{\omega \in {\cal B}}S(\sigma ^{\otimes n}||\omega )=$ $S(\sigma
^{\otimes n}||\omega _{n}).$ Thus we have that 
\[
D^{BE}\leq \frac{1}{n} \min_{\omega \in {\cal B}}S(\sigma ^{\otimes n}||\omega ) 
\]

\noindent and this completes our proof ${}_{\Box}$.

\bigskip

\noindent {\bf Corollary. }The bound entanglement assisted entanglement of
distillation of the state $\sigma $ is bounded from the above by the
relative entropy of entanglement of $\sigma .$

\noindent {\em Proof:} This follows immediately from the fact that $%
\min_{\omega \in {\cal B}}S(\sigma ^{\otimes n}||\omega _{n})\leq $ $%
n\min_{\omega \in {\cal B}}S(\sigma ||\omega _{1}).$ In particular, this
limit is also smaller then or equal to the entanglement of formation of the state $\sigma$ 
since the entanglement of formation is always greater than or equal to the relative entropy
of entanglement \cite{Vedral2}.

\bigskip \noindent Note that the above result is more general in that it
also applies if the assistance BE states are not all the same providing that
they have a positive partial transposition. This is because of the special
property of partial transposition, 
\[
(\beta_1\otimes \beta_2\otimes ... \otimes
\beta_n)^{T_2}=\beta_1^{T_2}\otimes \beta_2^{T_2}\otimes ... \otimes
\beta_n^{T_2} 
\]
where betas are all bi-partite states and $T_2$ is a partial transposition
over the second subsystem (the same is true if we partially transpose over
the first subsystem). However, if the BE states can have a negative partial
transposition, then the above might not in general apply as we discussed before.
The reason is that by "adding together" two of the BE states
with negative partial transposition, $\beta_1\otimes \beta_2$, we might
produce a free entanglement state. So, if we work under the assumption that 
a direct product of two BE states is a BE state, then, the above result 
applies even when the distillation is aided by different BE states.

Can the above discussed bound be improved? Suppose that $\Lambda$ is the LOCC
which distills $\sigma^{\otimes n}$ (for $n\rightarrow\infty$). Suppose also 
that we identify a set of states ${\cal G}$, such that $S(\psi^{\otimes m}||\omega_m)\ge m$
for all $\omega_m\in {\cal G}$ and $\Lambda(\omega_m)\in {\cal G}$. Then 
\[
\lim_{n\rightarrow\infty} \frac{\min_{\omega_n\in {\cal G}} S(\sigma^{\otimes n}||\omega_n)}{n}  
\]
is the upper bound on distillation. Of course, this might well be smaller than
the previous bound which was taking minimum over BE states only. However, this
kind of bound is not very useful since it is particularly tailored for $\Lambda$.
Once $\Lambda$ is known to distill $\sigma^{\otimes n}$, then we can also 
apply it directly and see what the efficiency is. This implies that if we are
to keep LOCC as general as possible, then the above bound can only be improved if we
find a different distance measure, $D$, to quantum relative entropy with the following properties: 1. $D(\Lambda(\sigma)||\Lambda(\rho))\le D(\sigma||\rho)$ for all $\Lambda$
which are LOCC and all $\sigma$ and $\rho$; 2. the minimum of $D$ over BE states for a maximally entangled pure state of two systems of dimension $n$ is $\log n$, just like in the case of the quantum relative entropy. The question of finding such a measure, if one should exist, remains an open problem.

\noindent Our result strongly suggests that bound entanglement cannot
enhance distillation in any way. In fact, Alice and Bob need to invest some
entanglement in order to create a BE state (although this might become arbitrarily 
small asymptotically), which than is not very useful in
distillation as the above result suggests. The above, however, does not
exactly prove this: it just shows that the upper bound on distillation
without the presence of BE is the same as when we have an arbitrary number
of BE systems present. So the above shows us that $D\leq D^{BE}\leq
\min_{\omega \in {\cal B}}S(\psi ^{\otimes n}||\omega )/n$, but we could still
have that $D<$ $D^{BE}.$ To strictly prove the equality we would have to
show that the above upper bound to distillation can be achieved, i.e. that $%
D=\min_{\omega \in {\cal B}}S(\psi ^{\otimes n}||\omega )/n.$ We leave this
task for future research.

\bigskip

\noindent {\bf Acknowledgments.} I would like to thank D. P. DiVincenzo and M.
Horodecki for discussions on the subject of bound entanglement and
its usefulness in general. I would also like to thank the IBM quantum
information group as well as C. Panos and the Aristotle University in
Thessaloniki on their hospitality during my visits when this work was
completed.

\end{document}